\documentclass{article}
\usepackage[utf8]{inputenc}

\usepackage{graphicx}

\usepackage{siunitx}
\usepackage{upgreek}
\usepackage[version=4]{mhchem} %chemical formula gas

\usepackage[square,numbers]{natbib}%use this citation style
\usepackage{authblk}

\title{Performance of a GridPix TPC readout based on the Timepix3 chip \\\vspace{3mm} \normalsize Talk presented at the International Workshop on Future Linear Colliders (LCWS2018), Arlington, Texas, 22-26 October 2018. C18-10-22 }

\author[1]{C. Ligtenberg}
\author[1,2]{K. Heijhoff}
\author[2]{Y. Bilevych}
\author[2]{K. Desch}
\author[1]{H. van der Graaf}
\author[2]{M. Gruber}
\author[1]{F. Hartjes}
\author[2]{J. Kaminski}
\author[1]{N. van der Kolk}
\author[1]{P.M. Kluit}
\author[1]{G. Raven}
\author[2]{L. Scharenberg}
\author[2]{T. Schiffer}
\author[2]{S. Schmidt}
\author[1]{J. Timmermans}

\date{24 October 2018}

\affil[1]{\normalsize Nikhef, Science Park 105, 1098 XG Amsterdam, The Netherlands}
\affil[2]{\normalsize Physikalisches Institut, University of Bonn, Nussallee 12, 53115 Bonn, Germany}

\begin{document}

\maketitle
\begin{abstract}
With the ultimate goal of developing a pixel-based readout for a TPC at the ILC, a
GridPix readout system consisting of one Timepix3 chip with an integrated amplification grid was embedded
in a prototype detector. The performance was studied in a testbeam with 2.5 GeV electrons at the
ELSA accelerator in Bonn. The error on the track position measurement both in the drift direction and in the readout plane is dominated by diffusion. Systematic uncertainties are limited to below \SI{10}{\um}. The GridPix can detect single ionization electrons with high efficiency, which allows for energy loss
measurements and particle identification. From a truncated sum, an energy loss (dE/dx) resolution
of 4.1\% is found for an effective track length of 1 m. Using the same type of chips, a Quad module was developed that can be tiled to cover a TPC readout plane at the ILC. Simulation studies show that a pixel readout can improve the momentum resolution of a TPC at the ILC by about \SI{20}{\percent}.
\end{abstract}

\section{Introduction}
For the International Large Detector (ILD) at the International Linear Collider, a gaseous Time Projection Chamber (TPC) is foreseen as the central tracker \cite{Behnke:2013xla}. The TPC provides a large number of measurement points, has a minimal material budget and can identify particles by energy loss (dE/dx) measurements. In the baseline design, the TPC has a pad readout in combination with GEM or Micromegas devices for gas amplification. Because the size of the pads is much larger than the ionization scale, the granularity is a limiting factor in energy loss measurements. A GridPix pixel readout has a much finer granularity and can extract all information from a track.

In this paper the progress towards developing a pixel-based readout for the ILD TPC will be presented. 

\section{A GridPix based on the Timepix3 chip}
A GridPix is a readout system consisting of a pixel readout chip and an integrated amplification grid coupled by microelectronic post-processing techniques \cite{Colas:2004ks, Kaminski:2017bgj}. The chip is a Timepix3 chip with \SI{256 x 256}{} pixels with a pixel pitch of \SI{55}{\um} \cite{Poikela:2014joi}. A grid is located \SI{50}{\um} above the chip and supported by SU8 pillars. The \SI{1}{\um} thick Aluminum grid has \SI{35}{\um} diameter circular holes aligned to the pixels. In order to prevent damage from discharges of the grid, the Timepix3 chip has a \SI{4}{\um} thick Silicon-rich Silicon Nitride protective layer.

A GridPix detects ionizing particles through efficient detection of all ionization electrons that are liberated in the gas. Ionisation electrons are drifted towards the readout by an electric field until they reach the amplification region. Here they cause an ionization avalanche that is collected on a single pixel pad. The Timepix3 chip pixels have low electronic noise ($\approx$70 e$^-$) and can register a precise time of arrival (ToA) using a \SI{640}{MHz} TDC. Simultaneously it can record the time over threshold (ToT) using a \SI{40}{MHz} clock. The data-driven readout is performed with a SPIDR board \cite{Visser:2015bsa}.

\section{Results from 2017 testbeam}
 A small detector with one Timepix3 based GridPix was tested in July 2017 using a test beam of 2.5 GeV electrons. The results have been previously published in \cite{Ligtenberg:2018sjs}. Here we will restrict ourselves to the most relevant results.

\subsection{Description of the GridPix detector}
\begin{figure}
    \centering
    \includegraphics[width=0.6\textwidth]{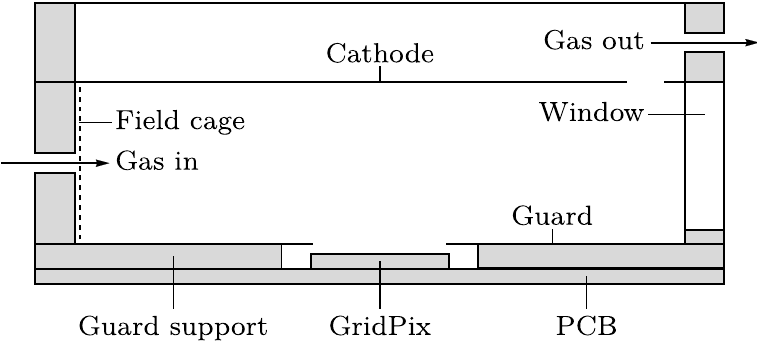}
    \caption{Schematic drawing of the GridPix detector.}
    \label{fig:detector}
\end{figure}

The Timepix3 based GridPix is embedded in a small drift volume as shown in figure \ref{fig:detector}. The dimensions of the box are \SI{69x42x28}{mm}. A cathode is located approximately \SI{20}{mm} above the GridPix. The electric field was kept homogeneous using a cage of conductive strips and a guard electrode at a height of \SI{1}{mm} above the GridPix. The volume was flushed with a gas mix of \SI{95}{\percent} \ce{Ar}, \SI{3}{\percent} \ce{CF4}, and \SI{2}{\percent} \ce{iC4H10} called T2K gas, a good candidate for the ILD TPC gas. The electric field was set to \SI{280}{V/cm}, near the value for which the gas reaches its predicted maximal drift velocity of \SI{78.86\pm0.01}{\um/ns} \cite{Biagi:1999nwa}. The hit $z$-position was calculated using the predicted drift velocity and the hit ToA. The grid voltage was \SI{350}{V} ensuring a high gain and consequently a high efficiency. The threshold was set to about \SI{800}{e^-} to reduce the noise to a minimum.

\subsection{Setup at test beam}

\begin{figure}
    \centering
    \includegraphics[width=0.8\textwidth]{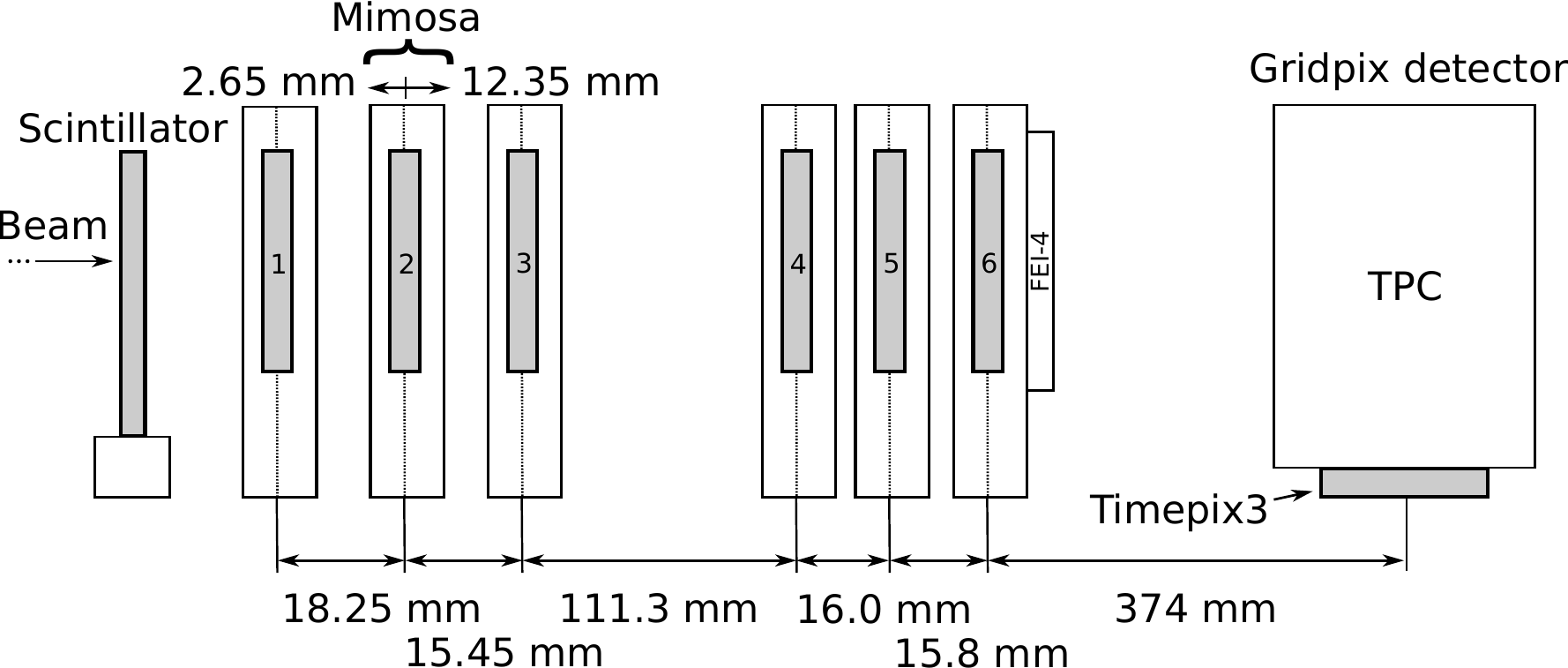}
    \caption{Setup with telescope and GridPix detector.}
    \label{fig:setup}
\end{figure}

The detector was probed with \SI{2.5}{GeV} electrons provided by the ELSA facility in Bonn at a maximum rate set to \SI{10}{kHz}. The setup is shown in figure \ref{fig:setup}. Electrons first passed through a scintillator used as a trigger and then through a Mimosa telescope with 6 silicon detector planes used to give a reference track. The beam entered the GridPix detector drift volume through a \SI{5}{mm} synthetic window. The Timepix3 hits were attributed to a single trigger by considering all hits within \SI{400}{ns} of a trigger.

\subsection{Reconstruction and selection}
To reconstruct a track, a straight line was fitted to the hits using a simple linear regression fit. Hits were assigned errors as given in section \ref{sec:hit resolution} in the directions perpendicular to the beam. Because of a multiple scattering at the last telescope plane, only the intercept at this plane was used as a reference. Some basic selections were made to ensure a clean track. Among the most stringent ones are a cut on events with less than 30 hits in the TPC and a hit ToT of at least \SI{0.15}{\us}.

\subsection{Time walk correction}
\begin{figure}
    \centering
    \includegraphics[width=0.6\textwidth]{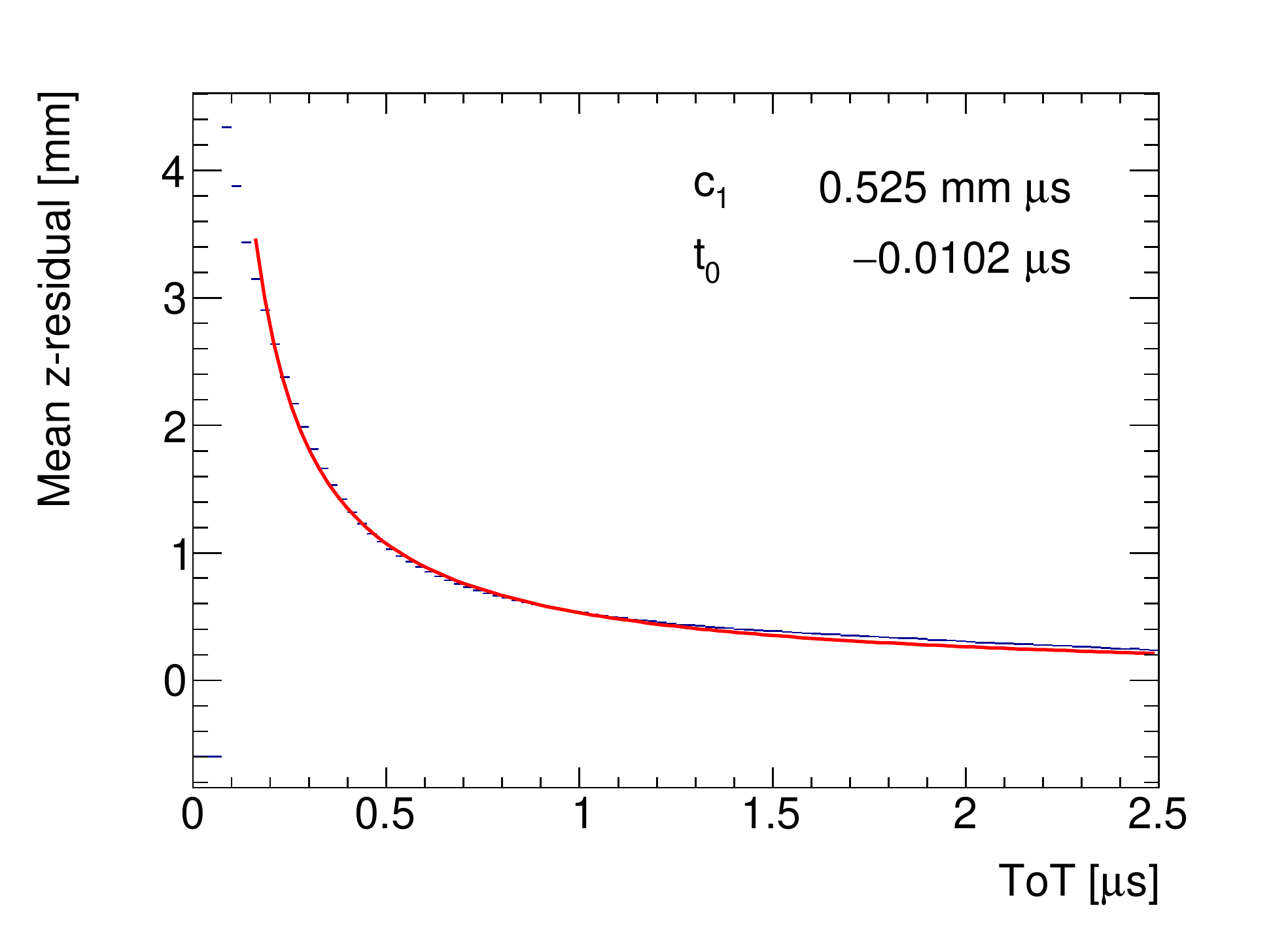}
    \caption{Mean track residual in the drift direction fitted with equation \eqref{eq:timewalk}.}
    \label{fig:timewalk}
\end{figure}

Time walk is caused by the dependence of the measured ToA on the magnitude of the signal. Using the ToT as measure of signal strength, the time walk can be corrected for, and the resolution can be improved. The correction can be parametrized using the time walk $\delta z_\text{tw}$ as a function of the corrected ToT $t_\text{ToT}$:
\begin{equation}
    \delta z_\text{tw} = \frac{c_1}{t_\text{ToT} + t_0},
    \label{eq:timewalk}
\end{equation}
where $c_1$ and $t_0$ are constants determined from a fit to the mean track residual in figure \ref{fig:timewalk}. The track residual is defined as the the difference between the hit position and the track fit prediction.

\subsection{Hit resolution} \label{sec:hit resolution}
\begin{figure}
    \centering
    \includegraphics[width=0.6\textwidth]{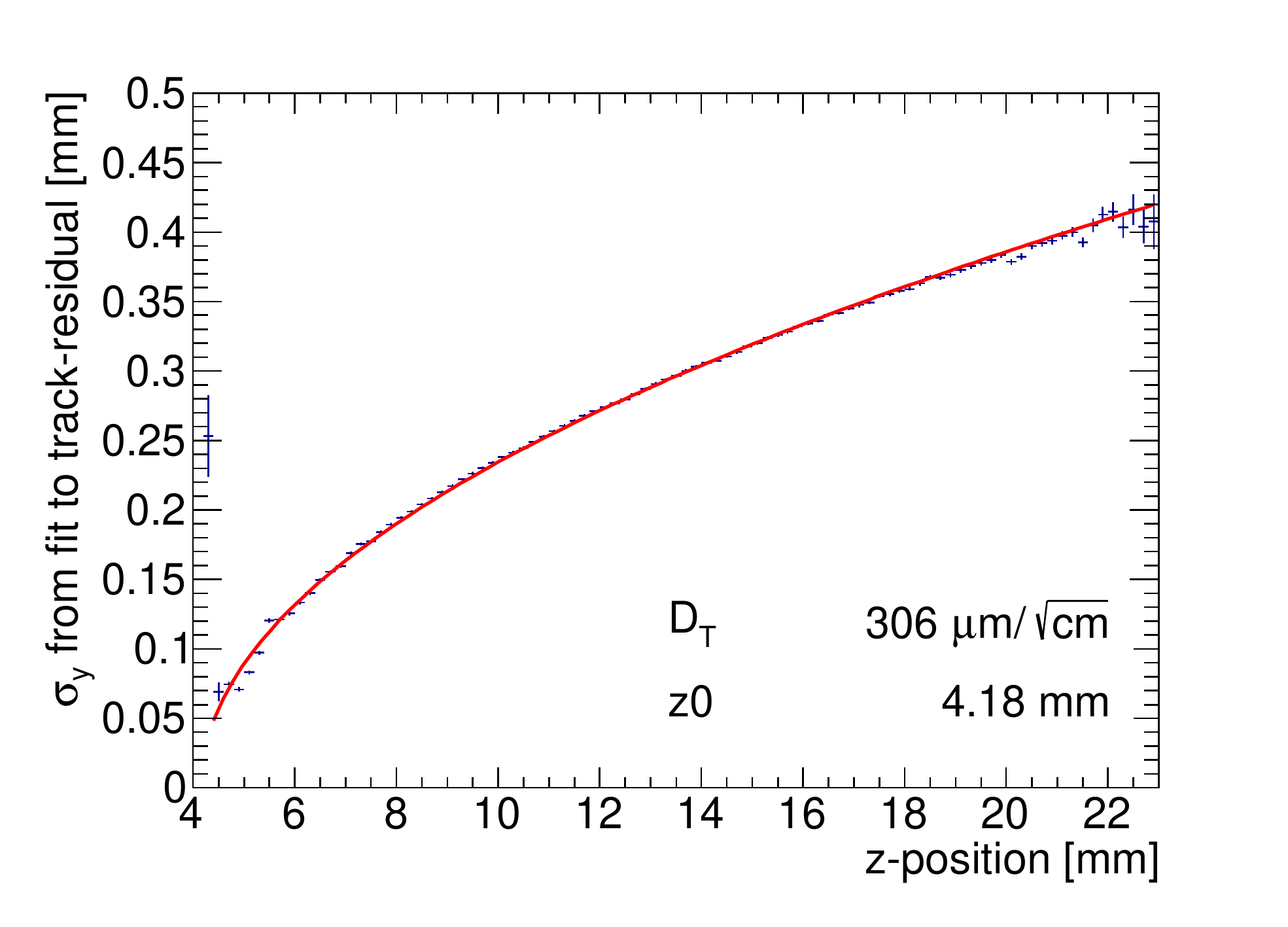}
    \caption{Measured hit resolution in pixel plane (blue points) fitted with the resolution according to equation \eqref{eq:sigmay} (red line), where the hit resolution at zero drift distance $d_\text{pixel}/\sqrt{12}$ was fixed to \SI{15.9}{\um}.}
    \label{fig:Diffy}
\end{figure}

The two main contributions to the hit resolution in the pixel plane are a constant contribution caused by the pixel size $d_\text{pixel}$ and a transverse drift component that scales with drift distance and the diffusion coefficient $D_T$. The resolution $\sigma_y$ is given by:
\begin{equation}
    \sigma_y^2=\frac{d_\text{pixel}^2}{12} +D_T^2(z-z_0),
    \label{eq:sigmay}
\end{equation}
where $z_0$ is the position of the grid. The hit resolution as a function of $z$-position is given in figure \ref{fig:Diffy}.

\begin{figure}
    \centering
    \includegraphics[width=0.6\textwidth]{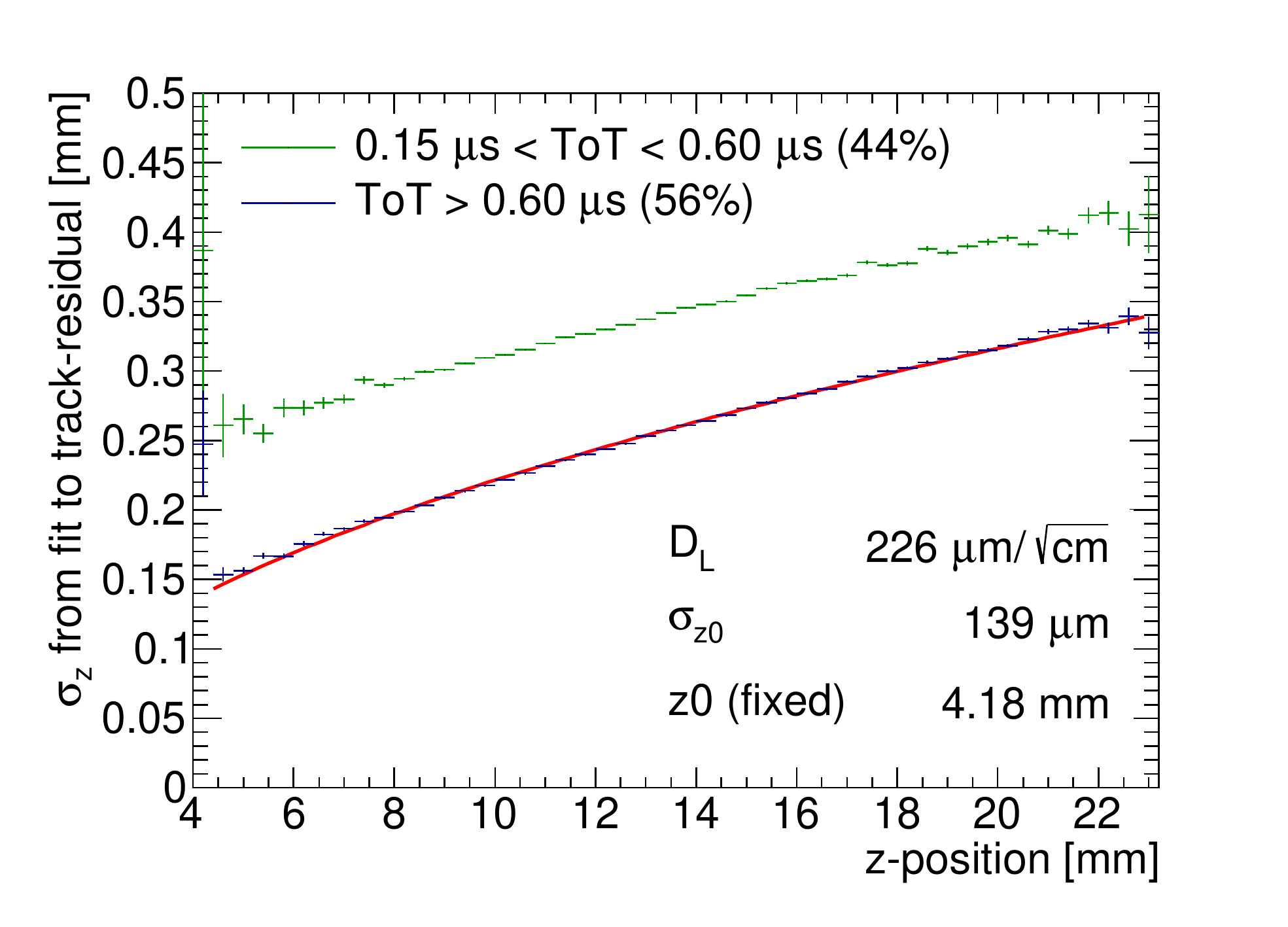}
    \caption{Measured hit resolution in drift direction split by ToT. The hits with a ToT above \SI{0.60}{\us} (blue points) are fitted with the resolution according to equation \eqref{eq:sigmaz} (red line). In the legend the fraction of hits in both selections is given.}
    \label{fig:Diffz}
\end{figure}

Likewise, the main contributions to the resolution in the drift direction are a constant contribution from the time resolution $\sigma_\tau$ of $\SI{1.56}{ns}$, a contribution from other noise sources such as jitter and time walk, and a contribution from longitudinal diffusion with coefficient $D_L$. The resolution $\sigma_z$ is given by:
\begin{equation}
    \sigma_z^2=\frac{\sigma_\tau^2 v_\text{drift}^2 }{12}+\sigma_{z0}^2+D_L^2(z-z_0).
    \label{eq:sigmaz}
\end{equation}
This resolution is shown in figure \ref{fig:Diffz}, where the hits with a ToT below \SI{0.60}{\us} were shown separately because of the large time walk error they have.

\subsection{Deformations}
\begin{figure}
    \centering
    \includegraphics[width=0.6\textwidth]{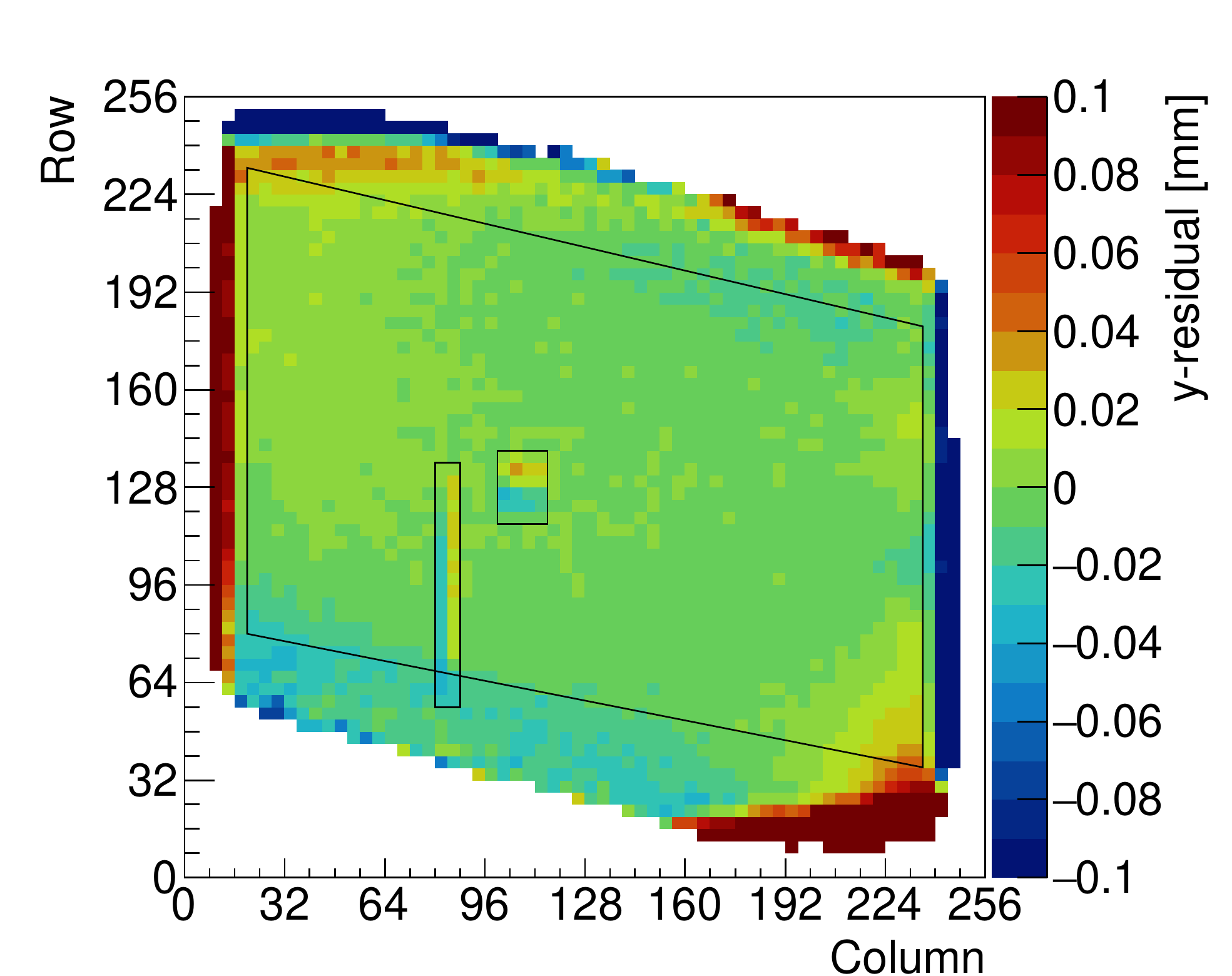}
    \caption{Mean residuals in the pixel plane at the expected hit position.}
    \label{fig:deformyexp}
\end{figure}

\begin{figure}
    \centering
    \includegraphics[width=0.6\textwidth]{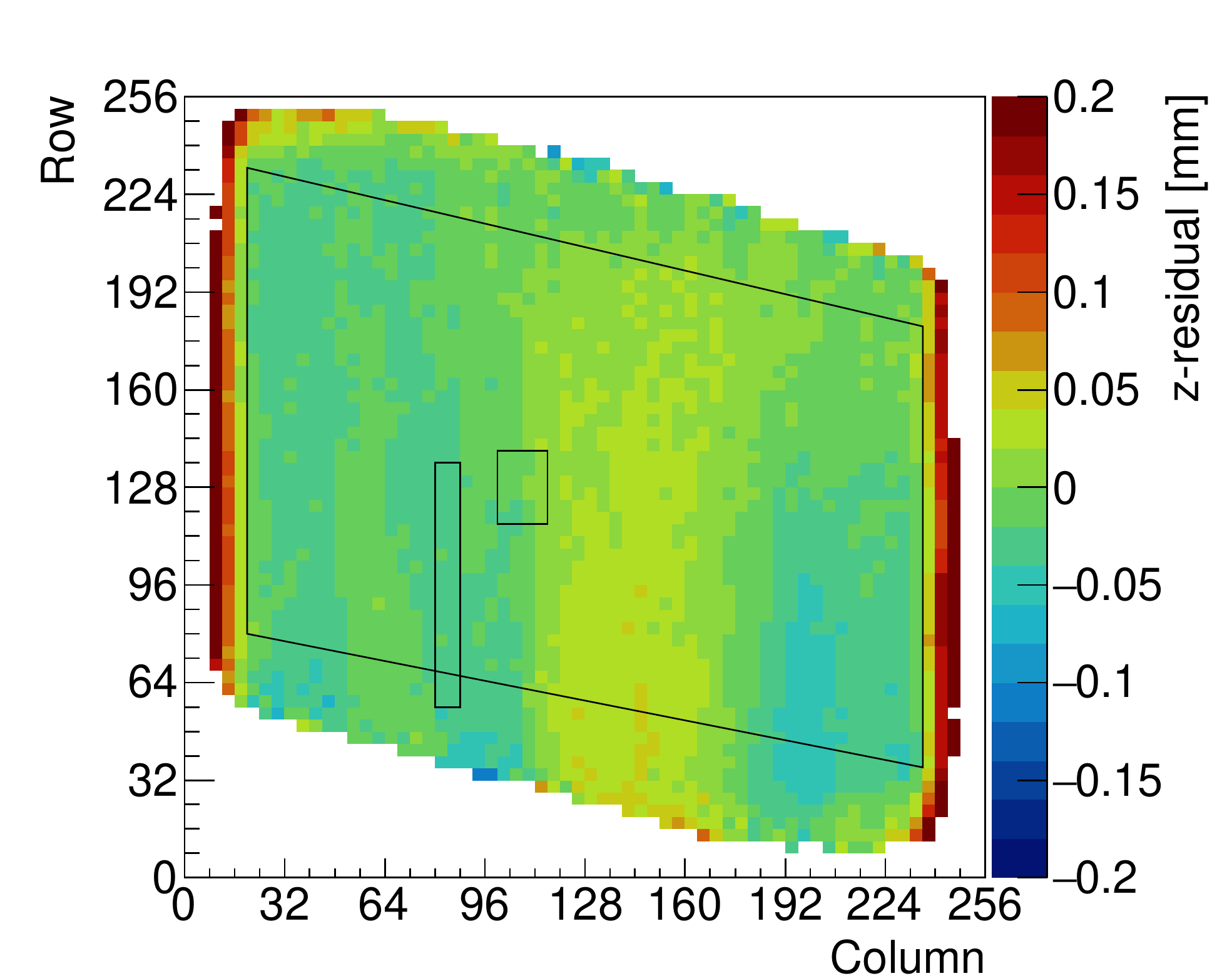}
    \caption{Mean residuals in the drift direction at the expected hit position.}
    \label{fig:deformxexp}
\end{figure}

Systematic deviations in the hit position measurements affect the performance of a large TPC and should typically be smaller than \SI{20}{\um} in the bending plane. In figure \ref{fig:deformyexp} and \ref{fig:deformxexp} the mean residuals in the plane, and in the drift direction are calculated for bins of \SI{4x4}{} pixels. In the fiducial area outlined with a black line, the systematic error given as the r.m.s. is \SI{8}{\um} in the pixel plane and \SI{31}{\um} (\SI{0.4}{ns}) in the drift direction.


\section{Particle identification by energy loss (dE/dx) measurements}
By measuring the characteristic energy loss (dE/dx) of a particle in the TPC its species can be identified. The performance here will be presented for an effective track length of \SI{1}{m} that is calculated by stringing 83 single chip tracks from the test beam data together. The energy loss from the data will be compared to the energy loss distribution of a Minimum Ionizing Particle (MIP), estimated by scaling the hit positions of the electron track by a factor 0.7, or effectively scaling a \SI{0.7}{m} electron track to a \SI{1.0}{m} MIP track.

A direct measure of the energy loss with a pixel readout is obtained by counting the number of ionization electrons. However, a few high energy deposits cause fluctuations in the mean. A more reliable estimate can be obtained by several methods. 

\begin{figure}
    \centering
    \includegraphics[width=0.6\textwidth]{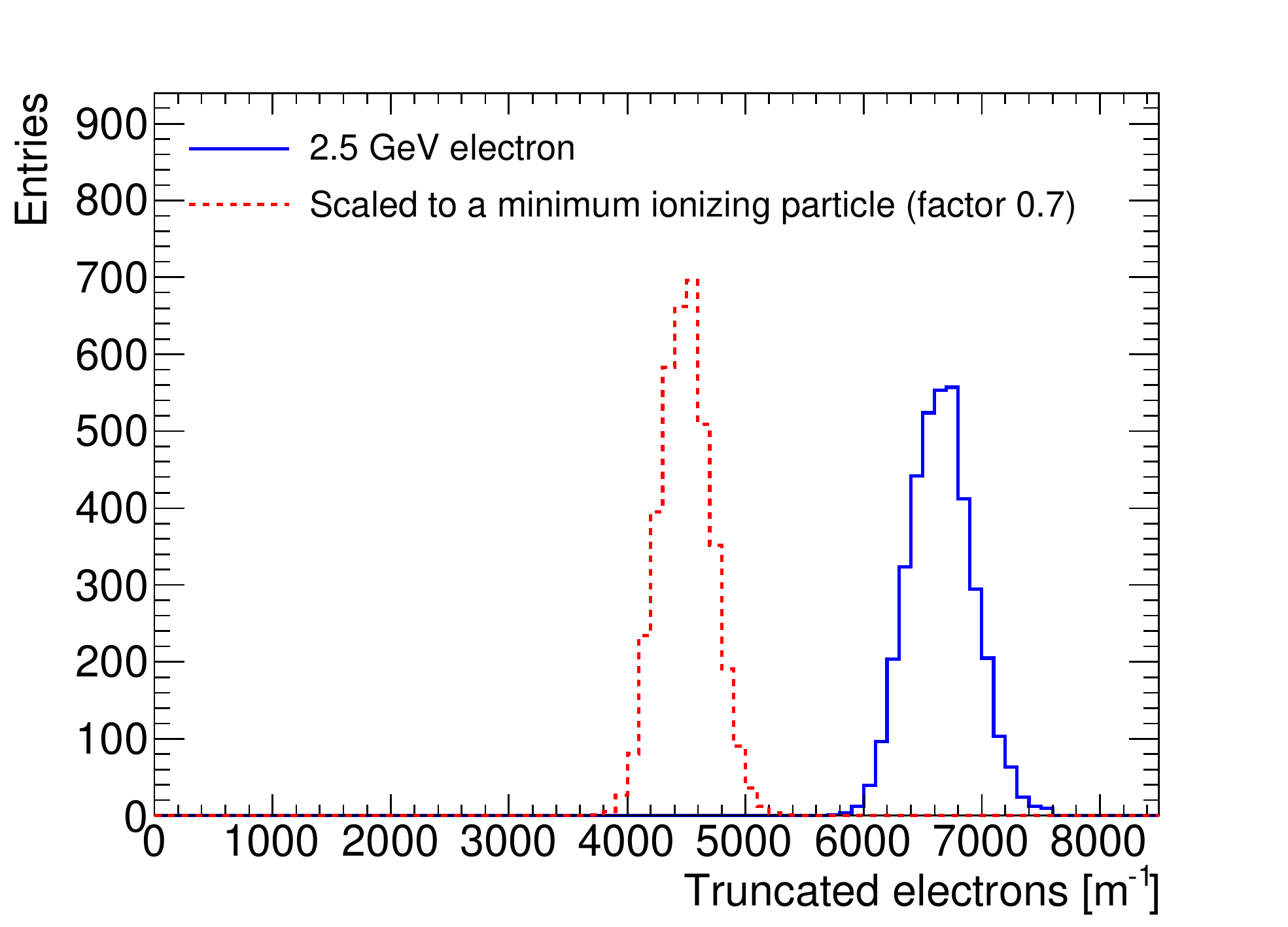}
    \caption{Distribution of truncated electrons per meter for the 2.5 GeV electron and the expected distribution for a minimum ionizing particle.}
    \label{fig:truncatedMean}
\end{figure}

One simple method that is also commonly adopted for TPC pad readouts is the truncated sum. For the GridPix readout it works in the following steps. First the number of electrons is summed for 20 pixel intervals. Secondly a fixed fraction of the intervals with the highest number of electrons is rejected. For the GridPix readout the best estimate was obtained by rejecting the top \SI{10}{\percent}. Finally, the other \SI{90}{\percent} is summed to retrieve a truncated sum. The result of this procedure for electron and MIP tracks of \SI{1}{\m} is shown in figure \ref{fig:truncatedMean}. The energy loss resolution defined as the standard deviation divided by the mean is \SI{4.1}{\percent} for a \SI{2.5}{GeV} electron.

The truncated mean uses slices of 20 pixels and does not make use of the fine granularity of the pixel readout. If clusters can be resolved, particle identification can be improved by counting the number of primary ionization clusters \cite{Hauschild:2002jh}.

\begin{figure}
    \centering
    \includegraphics[width=0.6\textwidth]{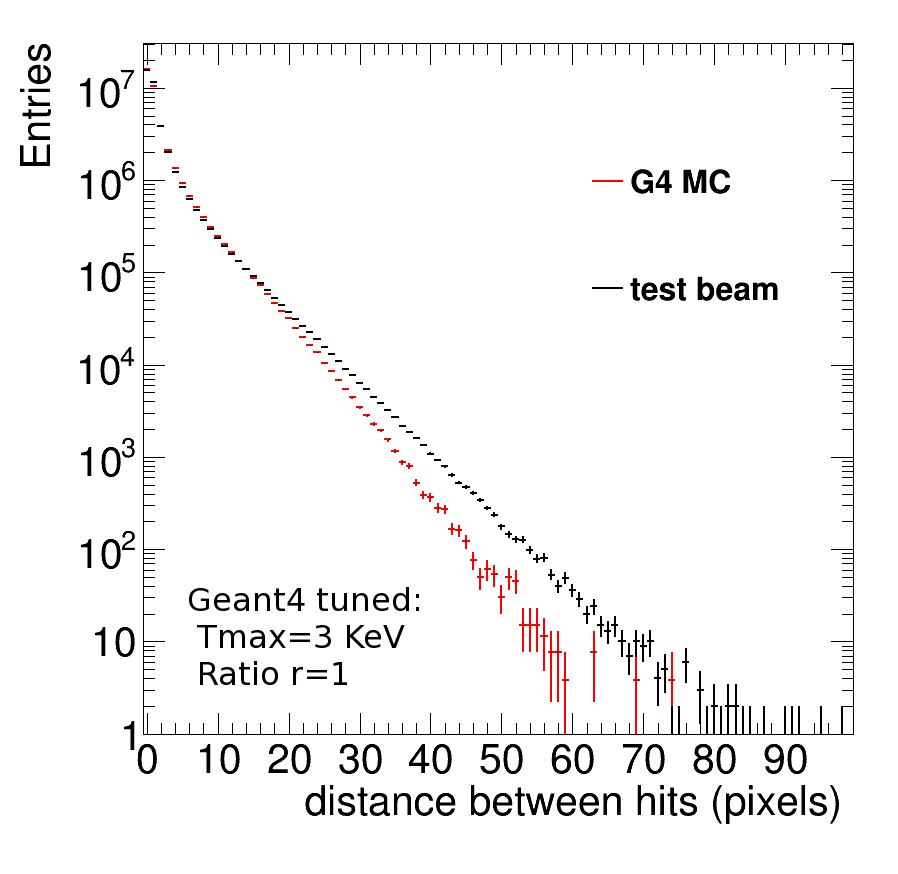}
    \caption{Mean distance between hits for a 2.5 GeV electron from test beam data and from a simulation.}
    \label{fig:meanDistance}
\end{figure}

A second method that does make use of the full granularity of the pixel detector is the calculation of a weighted mean distance between the pixel hits in the direction along the track. The mean distance distribution is shown in figure \ref{fig:meanDistance}. The mean distance is also calculated with data from a simulation. The simulation in GEANT4 has layers of gas with a thickness equal to the pixel pitch. In order to match the simulated data to our test beam data the parameters \texttt{Tmax} and \texttt{r} of the \texttt{G4UniversalFluctuation} model and the electron conversion threshold were tuned to \SI{3}{KeV}, 1 and \SI{27}{eV} respectively. The simulation serves as a cross-check for the data and is used to estimate the number of primary clusters, required as input to calculate the weights below.

The weighted mean for a track $\mu'$ is calculated using
\begin{equation}
    \mu' = \frac{1}{N_\text{hits}}\sum_{i=0}^{N_\text{hits}} w(d_i) d_i,
\end{equation}
where $N_\text{hits}$ is the total number of hits, $d_i$ is the distance between subsequent hits in the direction along the track and $w(d)$ is the weight as function of the distance. The assigned weights are the expected fluctuations from a pure Poisson distribution divided by the actual fluctuations. The effect is that hits at short distances, which are more likely to come from the same cluster, get a small weight, and hits at larger distances, which are more likely to come from separate clusters, get a larger weight.

\begin{figure}
    \centering
    \includegraphics[width=0.6\textwidth]{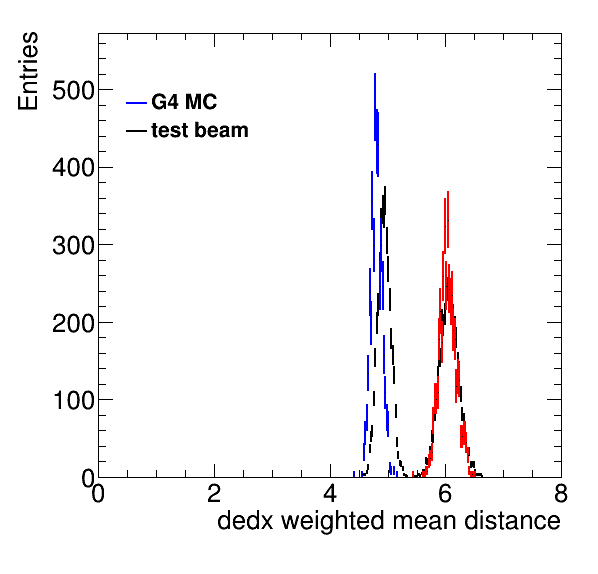}
    \caption{Weighted mean as described in the text for a 2.5 GeV electron from data and simulation and for a MIP from scaled data and simulation.}
    \label{fig:weightedMean}
\end{figure}

The weighted mean distance for an electron from data and simulation and a MIP from scaled data and simulation is shown in figure \ref{fig:weightedMean}. The resolution, again expressed as standard deviation divided by the mean, for an electron with this method is \SI{2.7}{\percent}. However, because the weighted mean distance is not proportional to the energy loss this is not most relevant measure.

How well the detector will be able to identify particles can be measured by the separation power $S$, here defined as
\begin{equation}
    S=\frac{\mu_e-\mu_\text{MIP}}{\sqrt{(\sigma_{e}^2+\sigma_\text{MIP}^2)/2}},
\end{equation}
where $\mu_e$ and $\mu_\text{MIP}$ are the mean (of the truncated sum or the weighted mean distance) for the electron and a MIP. $\sigma_e$ and $\sigma_\text{MIP}$ are the standard deviation (of the truncated sum or the weighted mean distance) for an electron and a MIP. The separation power $S$ for a \SI{1}{\m} long track of data is 8.8 using a truncated sum and 9.8 using the weighed mean distance.

\section{Quad module development}
\begin{figure}
    \centering
    \includegraphics[width=0.4\textwidth]{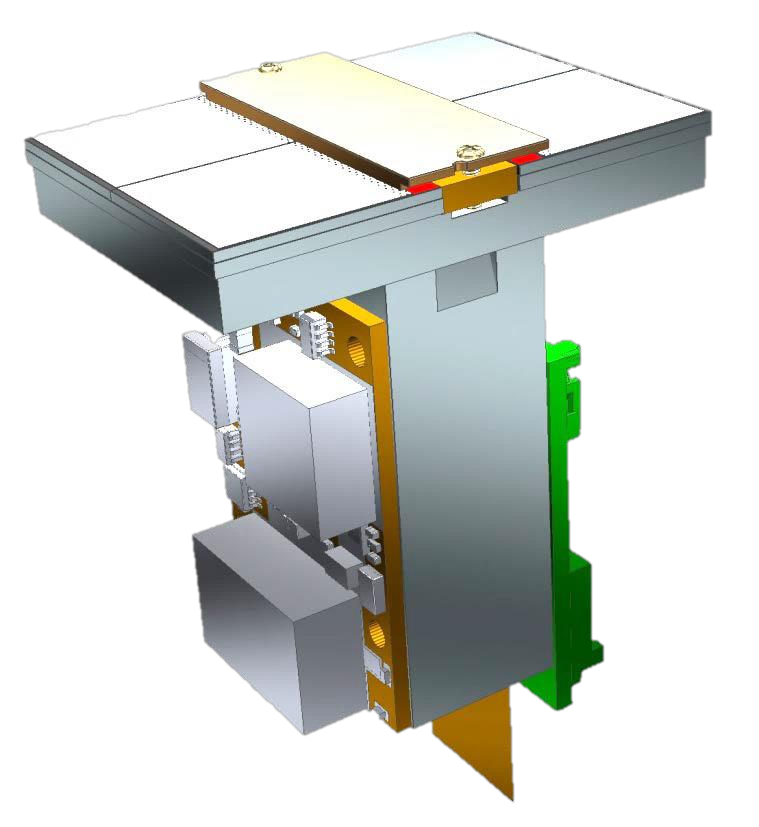}
    \caption{Rendered image of the Quad design}
    \label{fig:QuadRender}
\end{figure}

A four chip Quad module was developed with all services (low voltage regulator, IO-connections, and cooling) under the active surface. The Quad can be tiled to cover an arbitrary large area. In figure \ref{fig:QuadRender} a computer generated image of the Quad module is shown. Four chips are mounted on a plate that provides the cooling (cold carrier). All four chips are connected to one central wire bond PCB, that is covered with a guard electrode. The LV regulator is mounted on a hollow stump that is used to mount the Quad. 

The Quad has an active area coverage of \SI{69}{\percent}. A realistic tiling of the Quad module on the ILD TPC gives a coverage of \SI{59}{\percent}. With this coverage, the effective track length at \SI{90}{\degree} polar angle is \SI{0.78}{\m}. 

In the design of the Quad module special care has been taken to minimize distortions in the electric field. By simulations requirements were identified to keep the distortions below the \SI{100}{\um} at 5 pixels distance from the edge. The chip to chip distance must remain smaller than \SI{100}{\um}, larger distances for example on the wire bond side have to be bridged by a guard. The height of the guard above the chips must be precise at the \SI{20}{\um} level.
 
\begin{figure}
    \centering
    \includegraphics[width=0.6\textwidth]{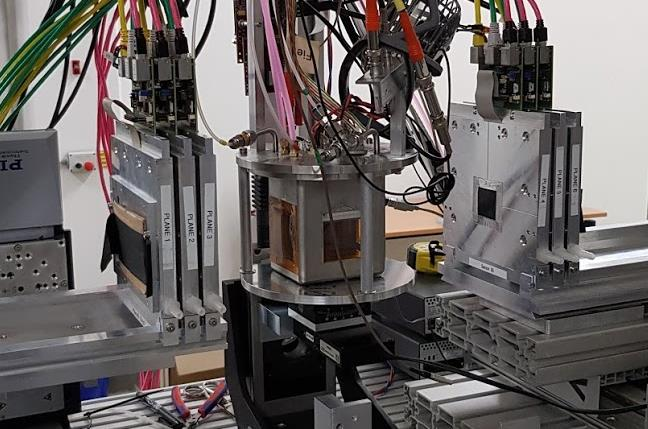}
    \caption{Picture of the Quad setup in the test beam at the ELSA facility in Bonn}
    \label{fig:QuadSetup}
\end{figure}

The first Quad modules have been produced, and in October 2018 two Quads were tested one by one at the ELSA facility in Bonn with \SI{2.5}{GeV} electrons. A picture of the setup is shown in figure \ref{fig:QuadSetup}. The Quad module is located inside the test box with the sensitive area facing downwards. The beam passes from left to right through three telescope planes, the Quad test box, and then another three telescope planes. Analysis of the data is ongoing.

\section{Simulation of ILD TPC with pixel readout}
In order to assess the performance of the ILD TPC with a pixel readout a full simulation  was made. As a starting point the ILD DD4HEP simulation (GEANT4) from ILCSoft was taken \cite{ILCSoft, Agostinelli:2002hh}. For single tracks, pixels were simulated by calculating the energy deposit in cylindrical shells with an active volume thickness equal to the pixel width of \SI{55}{\um}. In order to be able to simulate events with many tracks, a slightly larger granularity of \SI{990}{\um} with an interpolation step to \SI{55}{\um} was introduced. Tracks were reconstructed by a Kalman filter, which was also adapted to the pixel readout. From the reconstructed tracks the momentum resolution can be calculated.

\begin{figure}
    \centering
    \includegraphics[width=0.6\textwidth]{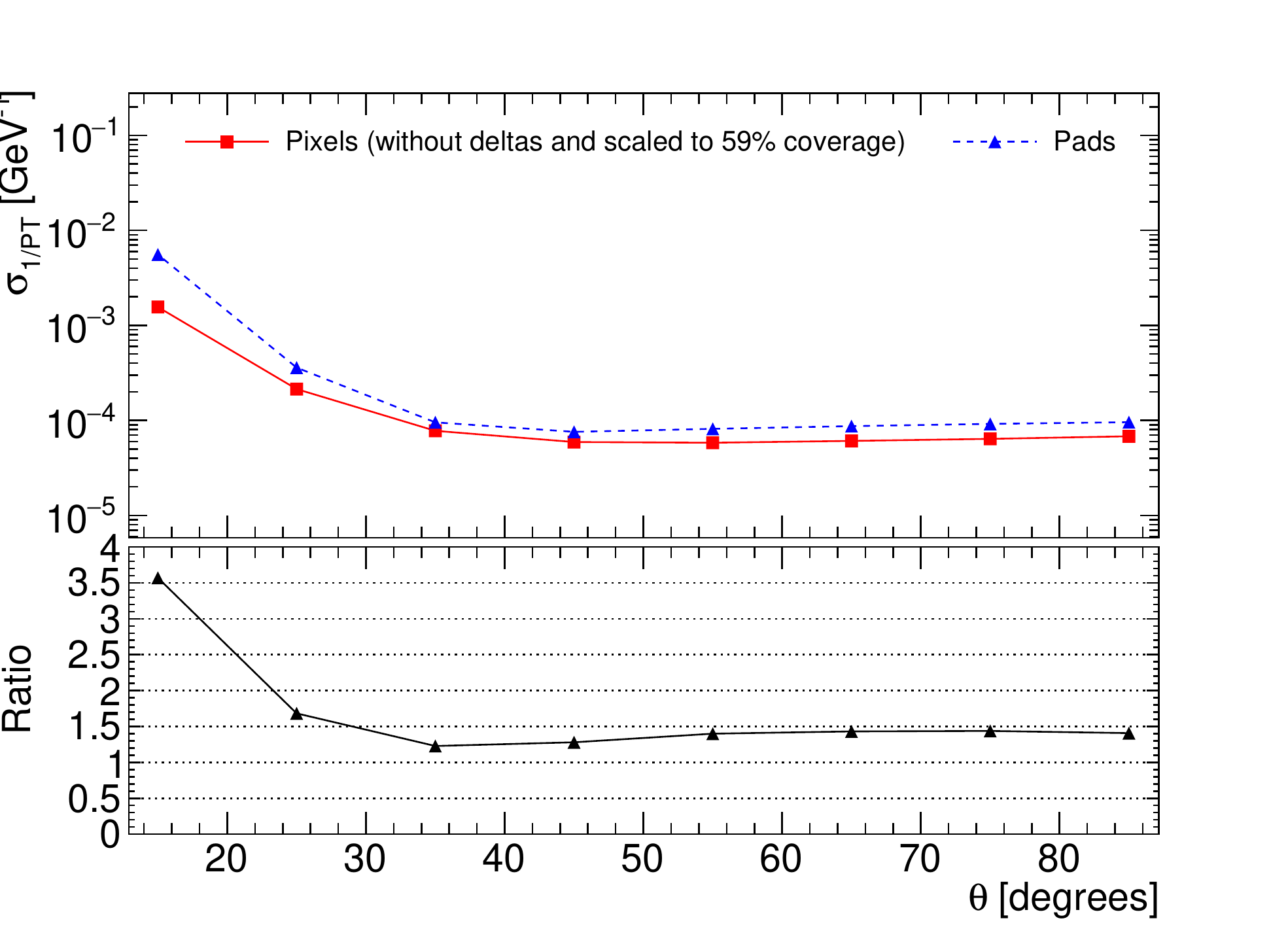}
    \caption{Momentum resolution ($\sigma_{1/P_T}$) of the TPC for a simulated 50 GeV muon for a pad and pixel readout}
    \label{fig:resolutionComparison}
\end{figure}

In figure \ref{fig:resolutionComparison} the momentum resolution of the TPC for a simulated 50 GeV muon is shown for a pad readout and for a pixel readout. At high momenta above 50 GeV the resolution is primarily limited by measurement errors. The pixel resolution is scaled from a simulation with \SI{100}{\percent} coverage to a realistic \SI{59}{\percent} coverage using a simple $\sqrt{N}$ factor. For all angles, the momentum resolution of a pixel readout is at least \SI{20}{\percent} better than for a pad readout. 
The difference is greater in the forward direction, because the number of hits for a pixel readout scales with the track length, whereas for a pad readout the number of hits is determined by the number of active pad rows which scales as the transverse component of the track length.

\section{Conclusions}
A GridPix based on the Timepix3 chip was reliably operated in a test beam setup. The resolution was found to be primarily limited by diffusion. Systematic uncertainties are smaller than \SI{10}{\um} in the pixel plane, which is small enough for a large TPC. The energy loss resolution (dE/dx) with a truncated sum is \SI{4.1}{\percent} per meter.  

In the next step towards a pixel readout for a TPC at the ILC, a four chip Quad module was developed that can be used to cover arbitrary large areas. Simulations show that a pixel readout for a TPC can give an improvement of at least \SI{20}{\percent} in moment resolution with respect to a pad readout. The Quad module was produced and has been tested in a test beam setup, from which results are expected soon.

\bibliography{mybibfile}

\begin{thebibliography}{10}
\providecommand{\natexlab}[1]{#1}
\providecommand{\url}[1]{\texttt{#1}}
\expandafter\ifx\csname urlstyle\endcsname\relax
  \providecommand{\doi}[1]{doi: #1}\else
  \providecommand{\doi}{doi: \begingroup \urlstyle{rm}\Url}\fi

\bibitem[Behnke et~al.(2013)Behnke, Brau, Foster, Fuster, Harrison, Paterson,
  Peskin, Stanitzki, Walker, and Yamamoto]{Behnke:2013xla}
Ties Behnke, James~E. Brau, Brian Foster, Juan Fuster, Mike Harrison,
  James~McEwan Paterson, Michael Peskin, Marcel Stanitzki, Nicholas Walker, and
  Hitoshi Yamamoto.
\newblock {The International Linear Collider Technical Design Report - Volume
  1: Executive Summary}, 2013.

\bibitem[Colas et~al.(2004)Colas, Colijn, Fornaini, Giomataris, van~der Graaf,
  Heijne, Llopart, Schmitz, Timmermans, and Visschers]{Colas:2004ks}
P.~Colas, A.~P. Colijn, A.~Fornaini, Y.~Giomataris, H.~van~der Graaf, E.~H.~M.
  Heijne, X.~Llopart, J.~Schmitz, Jan Timmermans, and J.~L. Visschers.
\newblock {The readout of a GEM- or micromegas-equipped TPC by means of the
  Medipix2 CMOS sensor as direct anode}.
\newblock \emph{Nucl. Instrum. Meth.}, A535:\penalty0 506--510, 2004.
\newblock \doi{10.1016/j.nima.2004.07.180}.

\bibitem[Kaminski et~al.(2017)Kaminski, Bilevych, Desch, Krieger, and
  Lupberger]{Kaminski:2017bgj}
J.~Kaminski, Y.~Bilevych, K.~Desch, C.~Krieger, and M.~Lupberger.
\newblock {GridPix detectors – introduction and applications}.
\newblock \emph{Nucl. Instrum. Meth.}, A845:\penalty0 233--235, 2017.
\newblock \doi{10.1016/j.nima.2016.05.134}.

\bibitem[Poikela et~al.(2014)Poikela, Plosila, Westerlund, Campbell, Gaspari,
  Llopart, Gromov, Kluit, van Beuzekom, FZappon, Zivkovic, Brezina, Desch, Fu,
  and Kruth]{Poikela:2014joi}
T~Poikela, J~Plosila, T~Westerlund, M~Campbell, M~De Gaspari, X~Llopart,
  V~Gromov, R~Kluit, M~van Beuzekom, FZappon, V~Zivkovic, C~Brezina, K~Desch,
  Y~Fu, and A~Kruth.
\newblock Timepix3: a 65k channel hybrid pixel readout chip with simultaneous
  toa/tot and sparse readout.
\newblock \emph{Journal of Instrumentation}, 9\penalty0 (05):\penalty0 C05013,
  2014.
\newblock URL \url{http://stacks.iop.org/1748-0221/9/i=05/a=C05013}.

\bibitem[Visser et~al.(2015)Visser, van Beuzekom, Boterenbrood, van~der
  Heijden, Muñoz, Kulis, Munneke, and Schreuder]{Visser:2015bsa}
J.~Visser, M.~van Beuzekom, Henk Boterenbrood, B.~van~der Heijden, J.~I.
  Muñoz, S.~Kulis, B.~Munneke, and F.~Schreuder.
\newblock {SPIDR: a read-out system for Medipix3 \& Timepix3}.
\newblock \emph{Journal of Instrumentation}, 10\penalty0 (12):\penalty0 C12028,
  2015.
\newblock \doi{10.1088/1748-0221/10/12/C12028}.

\bibitem[Ligtenberg et~al.(2018)]{Ligtenberg:2018sjs}
C.~Ligtenberg et~al.
\newblock {Performance of a GridPix detector based on the Timepix3 chip}.
\newblock \emph{Nucl. Instrum. Meth.}, A908:\penalty0 18--23, 2018.
\newblock \doi{10.1016/j.nima.2018.08.012}.

\bibitem[Biagi(1999)]{Biagi:1999nwa}
S.~F. Biagi.
\newblock {Monte Carlo simulation of electron drift and diffusion in counting
  gases under the influence of electric and magnetic fields}.
\newblock \emph{Nucl. Instrum. Meth.}, A421\penalty0 (1-2):\penalty0 234--240,
  1999.
\newblock \doi{10.1016/S0168-9002(98)01233-9}.

\bibitem[Hauschild(2002)]{Hauschild:2002jh}
M.~Hauschild.
\newblock {2D and 3D cluster counting with GEMs and small pads: The digital
  TPC?}
\newblock In \emph{{Linear colliders. Proceedings, International Workshop on
  physics and experiments with future electron-positron linear colliders, LCWS
  2002, Seogwipo, Jeju Island, Korea, August 26-30}}, pages 464--469, 2002.

\bibitem[ILC(2018)]{ILCSoft}
{ILCSoft}, 2018.
\newblock URL \url{https://github.com/iLCSoft}.

\bibitem[Agostinelli et~al.(2003)]{Agostinelli:2002hh}
S.~Agostinelli et~al.
\newblock {GEANT4: A Simulation toolkit}.
\newblock \emph{Nucl. Instrum. Meth.}, A506:\penalty0 250--303, 2003.
\newblock \doi{10.1016/S0168-9002(03)01368-8}.

\end{thebibliography}

\end{document}